\documentclass[useAMS,usenatbib,fleqn]{mn2e}

\usepackage{aas_macros}

\usepackage{epsfig}
\usepackage{epstopdf}
\usepackage{subfigure}
\usepackage{amsmath}    
\usepackage{longtable}
\usepackage{dcolumn}
\usepackage{amssymb}

\usepackage{wrapfig}

\DeclareGraphicsExtensions{.jpg,.eps,.EPS,.png,.pdf,.ps}

\usepackage{colordvi}
\usepackage{ulem}

\RequirePackage[colorlinks=true
,urlcolor=blue
,anchorcolor=blue
,citecolor=blue
,filecolor=blue
,linkcolor=blue
,menucolor=blue
,pagecolor=blue
,linktocpage=true
,pdfproducer=medialab
]{hyperref}

\usepackage[usenames,dvipsnames]{xcolor}

\title{Constraining MOND using the vertical motion of stars in the solar neighborhood}

\author[B. Margalit \& N. J. Shaviv]{
Ben Margalit$^{1,2}$\thanks{ben.margalit@mail.huji.ac.il} \& Nir J. Shaviv$^{1,3}$ \\ 
$^{1}$Racah Institute of Physics, Hebrew University of Jerusalem, Israel \\
$^{2}$Physics Department, Columbia University, 538 West 120th Street, New York, NY 10027 \\
$^{3}$Institute for Advanced Study, Einstein Drive, Princeton NJ, 08540, USA \\
}

\date{}

\begin{document}
\maketitle

\begin{abstract}
Stars with a different vertical motion relative to the galactic disk
have a different average acceleration. According to Modified Newtonian Dynamics (MOND) theories they should 
therefore have a different average orbital velocity while revolving around the Milky Way. 
We show that this property can be used to constrain MOND theories by studying stars in the local neighborhood.
With the {\sc hipparcos} dataset we can only place marginal constraints. However, the forthcoming {\sc gaia} catalogue with
its significantly fainter cutoff should allow placing a stringent constraint.  
\end{abstract}

\begin{keywords}
Galaxy: kinematics and dynamics, Galaxy: disc, solar neighbourhood
\end{keywords}

\section{Introduction}

The general inconsistency between observed luminous matter and the larger gravitationally inferred mass has led to the postulated existence of dark matter. An alternative explanation was proposed by \cite{Milgrom1983}, by modifying the Newtonian dynamics (MOND).

According to the standard non-relativistic formulation of MOND \citep{Milgrom1983}, a (Newtonian) force ${\bf F}$ acting on a body will cause it to accelerate according to the modified second law, $ {\bf F}=m\mu\left({|{\bf a}|}/{a_{0}}\right){\bf a}. $
Here $\mu(x)$ is the MONDian interpolating function, which behaves as
$\mu(x\gg1)\approx1$ and $\mu(x\ll1)\approx x$, while $a_{0}$ is a
global parameter with dimensions of acceleration. In order to explain
the observed flat rotation curves of galaxies, $a_{0}$ must be of
the order of magnitude of the galactic centripetal acceleration, i.e.
$a_{0}\sim 10^{-8} ~\text{cm} ~\text{s}^{-2}$. Two common choices for the interpolating function are the ``simple" function $\mu(x)=x/(1+x)$, and the ``standard" function $\mu(x)=x/\sqrt{1+x^{2}}$.

This simplest description (which does not obey the expected conservation laws) was reformulated by \cite{Bekenstein1984} using a modified Poisson equation,  with a non-relativistic MONDian gravitational potential.  
The resulting non-linear Poisson equation is 
\begin{equation} \label{eq:MOND_potential}
\nabla \cdot \left[ \mu \left( \left| \nabla \phi\right|/a_0\right) \nabla \phi\right] = 4 \pi G \rho.
\end{equation}

An alternative non-relativistic description is that of Quasilinear-MOND \citep[QuMOND,][]{Milgrom2010}, in which two potentials are defined. One is an auxiliary potential which solves the linear Poisson equation, and a second from which the force is derived. Respectively, they are 
\begin{equation} \label{eq:QUMOND_potential}
\Delta \phi_N = 4 \pi G \rho, ~~~  \Delta \phi = \nabla \cdot \left[ \nu \left( \left| \nabla \phi_N\right|/a_0\right) \nabla \phi_N\right].
\end{equation}
Here $\nu(y)$ is the inverted interpolating function, which is related to $\mu(x)$ through $\nu(y)=1/\mu[x(y)]$, where $y=x\mu(x)$. In the Newtonian regime $\nu(y \gg 1) \approx 1$, whereas in the deep MOND regime we have $\nu(y \ll 1) \approx y^{-1/2}$. For example, the inversion of the  standard interpolating function is $\nu(y) = \left( 1/2 + (y^{-2} + 1/4)^{1/2} \right)^{1/2} $. 

Within this description, the MONDian gravitational field ${\bf g} \equiv -\nabla \phi$ can be expressed in terms of the Newtonian field  
${\bf f} \equiv -\nabla \phi_N$ as 
\begin{equation} \label{eq:MOND_field}
{\bf g} = \nu \left( {f}/{a_0} \right) {\bf f} + {\bf \sigma}.
\end{equation}

This equation is derived immediately from eq.\ \ref{eq:QUMOND_potential}, where ${\bf \sigma}$ is a curl field ($\nabla \cdot {\bf \sigma} = 0$) that is generally non-zero and is defined so that $\nabla \times {\bf g} = 0$, as required for a conservative (irrotational) field\footnote{This can be more intuitively understood by considering ${\bf \sigma}$ as a ``magnetic" field generated by an effective current density 
${\bf j_{eff}} = (c /4 \pi)  (\nu^{\backprime}/ a_o) {\bf f} \times \nabla \left| {\bf f} \right| $.}. 
It is common to use eq.\ \ref{eq:MOND_field} while neglecting the curl field because this leads to a simple algebraic relation between the MONDian and Newtonian fields, and it does not involve solving a non-linear differential equation. Indeed, \cite{BradaMilgrom1995} have shown that such a curl field is usually small in comparison to the Newtonian field. In our present work we will be interested mostly in the vertical derivatives of the gravitational field, and it is not straightforward a priori that these derivatives are negligible. For this reason, we use the full form of eq.\  \ref{eq:MOND_field} in our analysis.
  
The non-relativistic description of MOND was also generalized \citep{Bekenstein2004} and written as a Lorentz-covariant theory of gravity, called the tensor-vector-scalar (TeVeS) theory. We will not discuss relativistic aspects, as they are irrelevant for the present work.
 
In addition to the study of MOND in the context of galactic rotation curves, which were the trigger for the development of MOND, MOND theories were studied and tested in various additional settings. These include globular clusters \citep{Scarpa2011,Ibata2011}, dwarf spheroidal galaxies \citep{Angus2008}, and galactic clusters, including both virialized and interacting \citep{PointecouteauSilk2005,Angus2007}. Our analysis will focus on testing MOND using the Milky-Way rotation curve in conjunction with the local stellar dynamics.
 
When treating stellar dynamics within the galaxy, it is important to remember that stars have a random component in addition to their galactic orbital motion, because of which they tend to oscillate radially and vertically about an orbiting guiding centre. Within the context of MOND, this additional motion offers the opportunity to locally sample stars experiencing a different average acceleration. The typical orbital acceleration at the solar galactocentric radius is of order $a_0$. The vertical acceleration is of order $a_z \sim \omega^2 z_\mathrm{max} \sim 3 \times 10^{-9} (\tau/60~\mathrm{Myr})^{-2} (z_\mathrm{max}/100~\mathrm{pc}) ~\mathrm{cm~s}^{-2}$ where $z_\mathrm{max}$, $\omega$ and $\tau$ are the amplitude, frequency, and period of the vertical oscillations. Thus, both vertical and orbital accelerations are comparable.
 
Within the simplest description of MOND (or its QuMOND formulation) the implications of this additional acceleration component are straightforward, as we will show in \S\ref{sec:MONDdrag}. Stars having a larger vertical amplitude should witness under MOND a smaller gravitational force, and therefore have a smaller average acceleration to remain in approximate circular orbits. 
This would then generate a systematic drift, linking the vertical amplitude with the extent of this local velocity drift. We use this fact to develop a test of MONDian dynamics in \S\ref{sec:MONDtest}. In \S\ref{sec:AstroData} we discuss the {\sc hipparcos} astrometric data set that we use for our analysis, and in \S\ref{sec:HipparcosApp} search for this effect in the presently available data, but do not find a vertically dependent drift. This is then used to place actual constraints on the MOND parameter space in  \S\ref{sec:MONDconstraints}. We then discuss the limitations of  this method in \S\ref{sec:discussion}.

\section{An average MONDian ``drag'' from the vertical acceleration of stars}
\label{sec:MONDdrag}

The essential point of the modified dynamics that we use in our analysis is directly apparent in the non-potential formalism of MOND. Only the absolute magnitude of the acceleration enters as an argument of the interpolating function. This feature effectively generates a coupling between the different axes of motion of a particle, and can in principle be used to distinguish between Modified and Newtonian dynamics. We begin by showing that this behaviour is a feature of the potential formalisms as well. 
Throughout the rest of this paper we adopt a Galactocentric cylindrical coordinate system $(r,\theta,z)$.

In addition to their circular motion around the galactic centre, stars exhibit vertical and radial motions about the galactic plane and their guiding radius respectively. For kinematically cool disk stars, this motion is well described by the epicyclic approximation \citep[see][chapter 3.2]{BinneyTremaine}, in which it is assumed that the vertical and radial oscillations are harmonic.

The MONDian drag effect can be qualitatively understood by comparing the radial (centripetal) acceleration of two hypothetical stars revolving in circular orbits about the galactic centre. The first resides in the galactic plane ($z=0$), while the second exhibits a vertical motion about the plane. Whereas the first star will experience a constant acceleration $a_r$ throughout its motion, the second star will experience a larger average net acceleration $|{\bf a}|=\sqrt{a_{r}^2+\langle a_{z}^2 \rangle}>a_{r}$. Following the MONDian argument, at smaller accelerations the ``effective" force that a Newtonian dynamicist would infer, $F / \mu(|{\bf a}|/a_0)$, is increased. Therefore, the first star will exhibit a larger centripetal acceleration than its vertically oscillating counterpart. This azimuthal drift of vertically oscillating stars relative to the non-oscillating baseline will increase with larger vertical accelerations $a_{z}$, and hence with greater vertical amplitudes $z_\mathrm{max}$.

We can find an analytic estimate for the effect $\Delta V(z_\mathrm{max})$ using perturbation theory. Assuming a circular orbit for the guiding centre, and expanding the radial distance and circular velocity as: $R=R_{0}+\Delta R$, $V=V_0+\Delta V$ for a population with the same angular momentum, the two deviations are  to first order related through
\begin{equation}\label{deltaR}
\Delta R=-\frac{\Delta V}{V_0}R_0.
\end{equation}

We can now expand the radial acceleration $a_r$ in terms of the velocity drift $\Delta V$
\begin{equation}
a_r=-\frac{(V_0+ \Delta V)^2}{R_0 + \Delta R} \approx -\frac{V_0^2}{R_0} - \frac{3V_0}{R_0}\Delta V.
\end{equation}

Next, we can also expand the gravitational field ${\bf g}$ about the unperturbed guiding centre at $r=R_0$ and $z=0$, where the derivates are evaluated, giving
\begin{equation}\label{eq:gravitationalField}
g_r\left(R,z\right) \approx \left.g_r\right\vert_0 + \left( \partial_r g_r \right)_{0} \Delta R + \frac{1}{2} \left( \partial_z^2 g_r \right)_{0}  \left< z^2 \right>,
\end{equation}
where $\left< z^2 \right>$ is a time average evaluated over a vertical orbital cycle. 

By identifying that $\left.g_r\right\vert_0 = -V_0^2/R_0$ and using eq.\ \ref{deltaR}, we find once we compare it to the MONDian acceleration that
\begin{equation} \label{eq:delta_V_1}
\left( -\frac{3V_0^2}{R_0} + \left( \partial_r g_r \right)_{0} R_0 \right) \frac{\Delta V}{V_0} = \frac{1}{2} \left( \partial_z^2 g_r \right)_{0} \left< z^2 \right> .
\end{equation}
This equation provides the velocity drift $\Delta V$ of a star in terms of various derivatives of the MONDian gravitational field and the $\left< z^2 \right>$ average of the star. 

We continue by deriving explicit expressions for the MOND field terms, which can then be plugged into eq.\ \ref{eq:delta_V_1}. Working with the QuMOND formalism, these terms can be found by taking derivatives of the MOND field as defined by eq.\ \ref{eq:MOND_field}. We make use of the following useful relations:
\begin{eqnarray}  
\partial_i f &=& { 1 \over f}  
\left( {f_r \partial_i f_r + f_z \partial_i f_z  } 
\right),  \nonumber \\
\left. f_z \right\vert_{0} &=& \left( \partial_r f_z \right)_{0} = \left( \partial_z f_r \right)_{0} = 0,  \nonumber \\
\left. f \right\vert_{0} &=& \left. -f_r \right\vert_0 .
\end{eqnarray}

Taking the curl of eq.\ \ref{eq:MOND_field}  and using the fact that both ${\bf g}$ and ${\bf f}$ are irrotational fields, we find that
\begin{eqnarray}
\partial_r \sigma_z - \partial_z \sigma_r &=& 
\frac{\nu^{\backprime}}{a_0}
{1 \over f} \left[     
{f_r}^2 \partial_z f_r + f_rf_z \partial_z f_z
\right. \nonumber \\
& &  \left.
- f_rf_z \partial_r f_r - {f_z}^2 \partial_r f_z 
\right],
\end{eqnarray}
where $\nu^{\backprime}$ is the derivative of the interpolating function. 

Taking the vertical derivative of this equation at the unperturbed guiding centre, we find
\begin{eqnarray} 
\label{eq:vertical_derivative_1}
\left. \partial_z^2 \sigma_r \right|_{0} &=& - \nu^{\backprime} \frac{|f_r|}{a_0} \left[ \partial_z^2 f_r + \left( \partial_r f_r \right) \left( \partial_z f_z \right) \Big/ |f_r| \right]_{0} \nonumber \\
& &+  \left. \partial_z \partial_r \sigma_z \right\vert_{0}.
\end{eqnarray}

Using the poloidal quadrupole morphology of the curl field \citep[e.g., as visualized in fig.\ 2 of][]{BradaMilgrom1995}, we note that  $\left. \partial_z \partial_r \sigma_z \right\vert_0 = \left. \partial_r (\partial_z \sigma_z) \right\vert_0 > 0$. Since this term is positive, we can place a lower bound on $\left. \partial_z^2 g_r \right\vert_{0}$ by discarding the last term. We will show below that such a bound will prove useful, and that in any case the leading contribution to the vertical derivative of the gravitational field arises from the algebraic part of eq.\ \ref{eq:MOND_field}. This algebraic derivative is
\begin{eqnarray} \label{eq:vertical_derivative_2}
\partial_z^2 \left[ \nu(f/a_0) f_r \right]_{0} &=& \nu \partial_z^2 f_r 
+ \nu^{\backprime} \frac{|f_r|}{a_0} \times \\ \nonumber
& &  
\left[ \partial_z^2 f_r - {\left( \partial_z f_z \right)^2} \Big/ {|f_r|} \right]_0.
\end{eqnarray}

By combining eqs.\ \ref{eq:vertical_derivative_1} and \ref{eq:vertical_derivative_2}, we find a lower bound on the vertical derivative of the MONDian gravitational field:
\begin{equation} \label{eq:vertical_derivative_result}
\left( \partial_z^2 g_r \right)_{0} \gtrsim \nu \left( \partial_z^2 f_r \right) - \frac{\nu^{\backprime}}{a_0} \left[ \left( \partial_z f_z \right)^2 + \left( \partial_r f_r \right) \left( \partial_z f_z \right) \right].
\end{equation}
Here $\nu$ and $\nu^\backprime$ are evaluated at the guiding centre.

Plugging eq.\ \ref{eq:vertical_derivative_result} into eq.\ \ref{eq:delta_V_1}, we find an explicit solution to the velocity drift $\Delta V$ (which is negative) as a function of the Newtonian gravitational field ${\bf f}$, and the interpolating function $\nu$. It is 
\begin{equation} \label{deltaV_MOND}
\frac{\left| \Delta V \right|}{V_0} \gtrsim \frac{\nu \left( \partial_z^2 f_r \right) - {\nu^{\backprime}}/{a_0} \left[ \left( \partial_z f_z \right)^2 + \left( \partial_r f_r \right) \left( \partial_z f_z \right) \right]}{{3V_0^2}/{R_0} - R_0 \partial_r g_r} \frac{\left< z^2 \right>}{2}.
\end{equation}
The second and third terms in the numerator express the purely MONDian effect due to the additional average vertical acceleration (as discussed in the beginning of this section). On the other hand, the first term in the numerator and the second in the denominator, both of which increase the velocity drift, are due to the form of the galactic potential and they exist also in the Newtonian limit.

Within the Newtonian terms, $\partial_z^2 f_r$ represents the falloff of the radial field perpendicular to the galactic plane. It physically expresses the fact that a star will spend more time outside the galactic plane due to its vertical oscillations, and thus in an average weaker radial gravitational field. This in turn causes a drift in the rotational velocity.
 
Similarly, the term $\partial_r g_r$ describes the falloff of the field with increasing galactic radius. Notice that $g_r$ in this term is the total acceleration (or MONDian) field, and not the Newtonian gravitational field. This is useful because the term is kinematically well determined. Any velocity drift must be accompanied by an outward shift in radius (as given by eq.\ \ref{deltaR}), where the galactic potential is weaker, thus contributing further to the velocity drift.

It is important to note that both these effects are due merely to the galactic potential model, and exist irrespective of whether the motion in this potential is assumed to follow Modified or Newtonian dynamics. In particular, these effects should contribute to give a Newtonian velocity drift, which is given mathematically by taking the Newtonian limit of eq.\ \ref{deltaV_MOND}, i.e., $\nu=1, \nu^{\backprime}=0$. As we now show, for most physically reasonable parameters, the Newtonian drift is significantly smaller than the expected MONDian effect.

Eq.\ \ref{deltaV_MOND} can in principle be used to asses the velocity drift for any galactic mass model by simply evaluating the gravitational field derivatives in this equation for the respective model. While this may prove useful for certain studies, in our present work we wish to estimate the velocity drift as broadly as possible instead of focusing on specific galactic models. To this extent we will try to further simplify eq.\ \ref{deltaV_MOND}, relying on observational constraints and a few relatively well established assumptions regarding the local galactic field, predominantly the vertical harmonic and radial epicyclic approximation.

The leading term contributing to the velocity drift expresses the affect of an additional vertical acceleration component. Under the harmonic approximation, this term can be written as
\begin{equation}
\partial_z f_z \approx -\omega^2 / \nu,
\end{equation}
where $\omega$ is the vertical oscillation frequency.

The term in the denominator, which expresses the radial falloff of the the galactic field, is fairly well constrained by observations and can be written using the Oort constants A and B \citep{OortConstants} as
\begin{equation}
\partial_r g_r = \frac{V_0}{R_0}\left(\frac{V_0}{R_0} + 2(A+B)\right)\approx 9\times10^{-4} \text{Myr}^{-2}.
\end{equation}
This result coincides relatively well with the naive order of magnitude estimate 
$\partial_r g_r \sim -{g_r}/{R_d} \sim \Omega^2$, where $R_d$ is the disk scale length (typically in the range $2-4 ~ \text{kpc}$) and $\Omega$ is the galactic rotation frequency.

The vertical falloff of the galactic field (the first term in the numerator of eq.\ \ref{deltaV_MOND}) is less stringently constrained by observations, but using the fact that the gravitational field is irrotational, we can roughly asses its magnitude to be
\begin{equation}\label{verticalFalloff}
\partial_z^2 f_r = \partial_r \left( \partial_z f_z \right) \sim \frac{\omega^2}{R_d} \sim 10^{-6} ~\text{Myr}^{-2} ~\text{pc}^{-1}
\end{equation}

Using this order of magnitude estimate we can compare the contributions of the terms in the numerator of eq.\ \ref{deltaV_MOND}. We find that the second, purely MONDian, term is a factor of about $\sim {\omega^2 R_d}/{a_0} \gtrsim 5$ larger than the first. This leads to a significantly more prominent effect under Modified dynamics than with Newtonian dynamics. On the other hand, the last term (which too is MONDian) is smaller than the leading (MONDian) term by roughly a factor of $\sim {\omega^2}/{\Omega^2} \approx 12$. We will nevertheless retain this term in our analysis as it further tightens the constraints we will find.

We conclude this section by giving a complete expression for the velocity drift, as a function of $z_\mathrm{max}$, by averaging over the vertical oscillations (which are much faster than the galactic rotation) and taking $\left< z^2 \right> = z_\mathrm{max}^2/2$. We also take into account deviations from the harmonic vertical motion as these will dampen the velocity drift at high $z_\mathrm{max}$. This is achieved by interpolating between harmonic acceleration close to the galactic plane, and constant acceleration far from the plane in the following manner
\begin{equation}
\label{eq:nonlinearz}
\left< \omega^2 \right>_z = { \omega_0^2 \Big/  \sqrt{1 + \left(z_\mathrm{max}/\zeta\right)^2}}.
\end{equation}
We have here introduced $\zeta$, a parameter with dimensions of length, which can be interpreted as the disk scale height. The Newtonian vertical field derivative can then be expressed as $\left< \partial_z f_z \right> \approx -\left< \omega^2 \right> \big/ \nu$.

Finally, we rewrite $\nu$ and $\nu^{\backprime}$ in terms of $\mu$ and $\mu^{\backprime}$. This is convenient since the argument of the latter interpolating functions is the well constrained galactic acceleration $a_r=V_0^2/R_0$, whereas the argument of the former is the unknown Newtonian field. Substituting these results into eq.\ \ref{deltaV_MOND}, we arrive at a final expression for the velocity drift as a function of $z_\mathrm{max}$. It is
\begin{eqnarray} \label{deltaV_MOND_2}
\frac{\left| \Delta V  \right|}{V_0} &\hskip -0.5mm \gtrsim \hskip -0.5mm& 
\left[  
\frac{1}{\mu} \left( \partial_z^2 f_r \right) +
\frac{\mu^{\backprime}}{\mu + ({V_0^2}/{R_0 a_0}) \mu^{\backprime}} \frac{\omega_0^4 / a_0}{1 + \left(z_\mathrm{max}/\zeta\right)^2} 
\right. \nonumber   \\ & &
\left.  
+ \frac{\mu^{\backprime}}{\mu a_0} \frac{\omega_0^2}{\sqrt{1 + \left(z_\mathrm{max}/\zeta\right)^2}} \left( \frac{V_0^2}{R_0^2} + 2\left(A+B\right)\frac{V_0}{R_0} \right) 
\right] \nonumber \\ & &
\times {
z_\mathrm{max}^2
\over 
{8V_0^2}/{R_0} - 8 V_0 \left( A + B \right)
}.
\end{eqnarray}

In the following section we compare this analytic formula (or rather a variant of it excluding the curl field) against the results of numeric simulations and show that the results are in good agreement. We later use this expression in \S\ref{sec:MONDconstraints} to constrain the parameters so that $\Delta V$ is consistent with observations.

In developing our results we have assumed that the stellar population is kinematically cool such that it is well described by the radial epicyclic approximation. This is an important point, as our theoretical prediction pertains to these young disk stars alone. Halo stars are not described by our model, but should ``seem" to drift relative to the Local Standard of Rest (LSR) if included in a sample of analyzed stars. This extraneous effect would arise since halo stars have a much smaller (z-component) angular momentum than disk stars, and would appear to have a $V$ component drift velocity relative to the disk stars. It is therefore essential to remove any halo stars from the data set before searching for any velocity drifts. This is discussed further in \S\ref{sec:AstroData}.

\section{The Drag in a realistic potential}
\label{sec:realistic}
Eq.\ \ref{deltaV_MOND_2} was obtained under a simplified potential. In this section we numerically estimate the expected drag for a realistic Galactic potential, as well as verify our analytic results.

We simulate test particle motion in a static Galactic potential under both Newtonian and MONDian dynamics. We first simulate the motion under a specific, yet complete, Galactic mass model, so as to asses the general expected magnitude of the drift effect. We model the Galactic potential in this case by integrating the density distribution given by `Model1' in \cite{DehnenBinney1998}. For the MOND simulation we take only the baryonic contribution to the density distribution, while for the Newtonian simulation we include also the dark matter halo as described by the full model of \cite{DehnenBinney1998}. To obtain the MOND potential we use the purely algebraic relation of eq.\ \ref{eq:MOND_field} (while neglecting the ${\bf \sigma}$ curl field for computational simplicity).

We use the standard interpolating function, and obtain a self consistent value for $a_0$ by ``fitting" the MONDian rotation curve to its Newtonian counterpart. This method yields a value of $a_0 = 2.06 \times10^{-8} ~\text{cm}~\text{s}^{-2}$ for the MOND acceleration parameter. Although this value is slightly higher than the $1.2\times10^{-8}~\text{cm}~\text{s}^{-2}$ found from surveys of external galaxies  \citep{Begeman1991,Gentile2011}, it has already been pointed out that higher values of $a_0$ are required to fit the Milky Way rotation curve using the standard interpolating function \citep[e.g.,][]{FamaeyBinney2005}.

\begin{figure*}
\centerline{
\epsfig{file=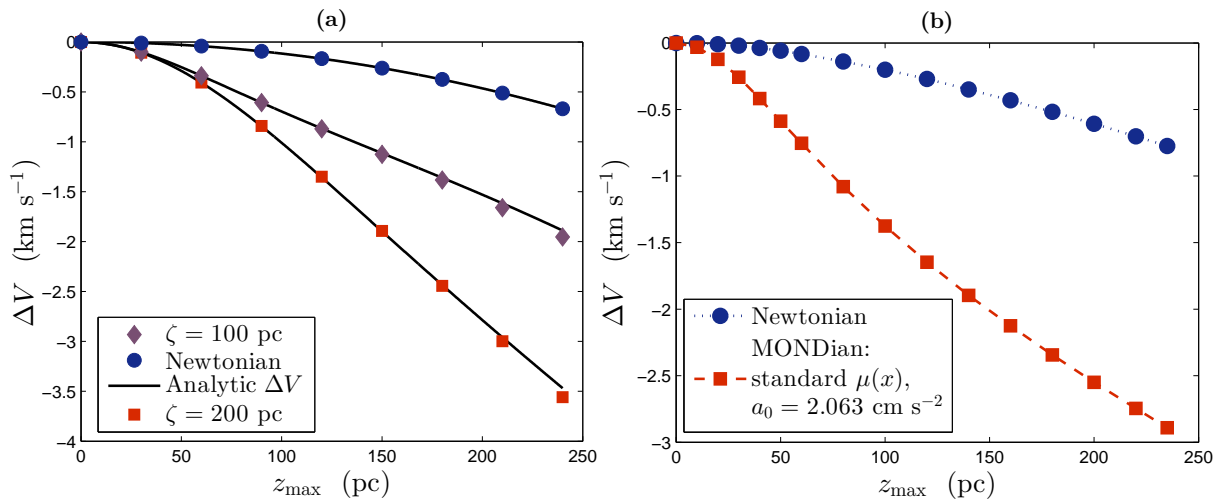, angle=0, width=0.95\textwidth}
}
\caption{The $V$ velocity drift as a function of the vertical oscillation amplitude $z_\mathrm{max}$. {\bf(a)} Simulation results (different symbols) and analytic approximation (solid lines) for the ``simplistic" galactic model described in the text. The analytic approximation is in good agreement with the simulation results, and begins to diverge slightly only at larger $z_\mathrm{max}$ as expected. The MONDian effect is characteristically several times larger than the expected Newtonian drift, but can be quenched at higher $z_\mathrm{max}$ by taking a small non-linearity scale $\zeta$. {\bf(b)} Simulation results for the galactic potential obtained by integrating mass model `Model1' \citep[from][]{DehnenBinney1998}. The results are similar in nature to those of panel {\bf(a)}. The velocity drift breaks from its quadratic dependence on $z_\mathrm{max}$ at around $\sim 100~\mathrm{pc}$, signifying a deviation from a purely harmonic vertical potential.
}
\label{fig:Combined_Vz} 
\end{figure*}

We then integrate the equations of motion for a series of test particles having the same initial Galactic radius $R_0$ and velocity $V_0$, but with different values of $z_\mathrm{max}$. This is achieved by implementing a 2D leapfrog integration scheme in the R-z plane. The radial field component is modified to account for rotation by adding the effective term $L_z^2/R^3$, where $L_z$ is the test particle's specific angular momentum.  For each run of the simulation, we extract the guiding radius velocity
$\left< V \right>$
about which the particle oscillates vertically and radially. The velocity drift $\Delta V$ is then found by subtracting 
$V_0$
 from each of these 
$\left< V \right>$
. The results comparing the Newtonian and MONDian velocity drifts obtained by these simulations are shown in figure \ref{fig:Combined_Vz} (b). It is clear from the figure that as expected in both cases the lag increases with larger $z_\mathrm{max}$, and that the effect is substantially more dramatic in the MOND case. Additionally, while the Newtonian lag is very nearly quadratic in $z_\mathrm{max}$, the MONDian effect breaks from this trend, due to non-Gaussianity in the vertical potential.

We also consider the simplified model discussed in obtaining Eq.\ \ref{deltaV_MOND_2}. This allows us to verify the validity of our analytic results, as well  as present the dependence on the various parameters. For the vertical field component we take a linear model at small $z_\mathrm{max}$, with the slope $\omega_0^2$, which evolves into a constant field at larger $z_\mathrm{max}$, exactly as presented in the previous section. For the radial component we take a linear dependence on radius and quadratic dependence on vertical displacement, giving us overall:
\begin{align} \label{eq:simplistic_galactic_model}
&f_z = -\frac{\omega_0^2}{ \sqrt{1+\left(z_\mathrm{max}/\zeta\right)^2} } z \nonumber \\
&f_r = {f_r}_0 + \left( \partial_r f_r \right)_0 \left(R-R_0\right) +  \frac{1}{2} \left( \partial_z^2 f_r \right)_0 z^2
\end{align}

We once again simulate MONDian and Newtonian test particle motion in this field and extract $\Delta V(z_\mathrm{max})$. We plot the numerical results as well as the analytic expression for the velocity drift (a revision of eq.\ \ref{deltaV_MOND_2} excluding the $\bf \sigma$ field contributions). The simulation results agree with the theoretical expression to within a few percent, with larger deviations from the analytic approximation at larger $z_\mathrm{max}$. The main source of deviation from eq.\ \ref{deltaV_MOND_2} is terms of higher order in ${\langle a_z^2 \rangle}/{a_r^2}$ which were neglected in deriving the result for sake of simplicity and coherence. Figure \ref{fig:Combined_Vz} (a) shows examples of the simulation results for a few different parameter values, as well as the respective analytic curves. One can notice from this figure the affect of choosing smaller $\zeta$ values---effectively quenching the additional vertical acceleration component, and hence the velocity drift, at higher $z_\mathrm{max}$. An additional way to decrease the velocity drift would be choosing very small values for the MOND acceleration parameter $a_0$, as in this limit we revert back to classical Newtonian dynamics.

\section{A test for MONDian dynamics}
\label{sec:MONDtest}

In previous sections we have presented the physical arguments and theoretical predictions of a MONDian velocity drift. In the following we propose using available astrometric data to plot the rotation velocity $V$ as a function of vertical oscillation amplitude $z_\mathrm{max}$ for a selection of stars in the solar neighborhood. By checking for a systematic lag $\Delta V(z_\mathrm{max})$ 
we attempt to constrain parameter values under MONDian dynamics.

We assume a harmonic vertical potential and assign a vertical amplitude to each star in our sample using the star's observables $W$ (vertical velocity) and $z$ (vertical displacement), that is
\begin{equation}\label{zmax}
z_\mathrm{max} = \sqrt{z^2+W^2/\omega_0^2}.
\end{equation}

The harmonic assumption is valid as long as we constrain ourselves to small vertical amplitudes, we therefore include anharmonic effects in our model as discussed in the previous section. \citet{Bahcall1985} show that the expected non-Gaussianity of the potential is $\sim15\%$ at $z_\mathrm{max}=100~\text{pc}$.

\subsection{The astrometric data}
\label{sec:AstroData}

We use the Extended Hipparcos Compilation (XHIP) catalog \citep{XHIP2012} in our analysis. This recent catalog is based on the New Reduction of Hipparcos data \citep{HipReduction} which includes parallax and proper motion entries for 117,955 stars in the solar neighborhood. The catalog also uses proper motions from the Tycho-2 Catalog \citep{Tycho2} in cases where the Hipparcos proper motions exceed Tycho error bounds by some extent. Additionaly, the catalog assigns radial velocities to 46,392 stars in the sample by cross referencing 47 sources of radial velocity data \citep[e.g.][]{PCRV}. We restrict our sample to kinematically complete well localized stars by taking only stars for which both proper motions and radial velocity exist, and by imposing a cutoff for stars with a quoted parallax error greater than 20\%. Thus we are left with a total of 32,418 stars in our sample.

In order to avoid possible contamination by halo stars we exclude relatively old stars for which $B-V < 0.68 \text{mag}$. We additionally exclude high velocity outliers by restricting our sample to stars within the velocity ellipsoid $({U}/{200})^2 + ({V}/{80})^2 + ({W}/{120})^2 < 1$. The constraints on $U$ and $W$ are quite loose for this cutoff ($\sigma_U \approx 25 ~{\mathrm{km}}~{\mathrm{s}^{-1}}$ and $\sigma_W \approx 11 ~{\mathrm{km}}~{\mathrm{s}^{-1}}$), yet the limit on $W$ is essentially immaterial as our method explicitly uses only specific ranges of $W$ (in the process of calculating $z_\mathrm{max}$) which are well below this limit, and the cutoff on $U$ was found to have little significance on the final results. The $V$ cutoff on the other hand is more stringent but is still well justifiable,  since $\sigma_V \approx 13-17 ~{\mathrm{km}}~{\mathrm{s}^{-1}}$. 
Throughout the analysis we take the LSR as $(U,V,W)_\mathrm{LSR}=(7.5,13.5,6.8) ~{\mathrm{km}}~{\mathrm{s}^{-1}}$ \citep{LSR}.

\subsection{Application of the Method to the Extended Hipparcos dataset}
\label{sec:HipparcosApp}

We focus now on applying our method to the sample discussed above.
A non-trivial task in performing this analysis is applying the statistics to an unbiased population. Since the number density of stars increases with decreasing galactic radii, we generally expect to find a non-Gaussian distribution of $V$ exhibiting an asymmetric drift. This artifact can be remedied by using only stars with a particular galactic angular momentum $L_z$.
Since a given value of $L_z$ corresponds to a particular guiding radius (for a population with some $z_\mathrm{max}$), stars with the same angular momentum will be observed in different phases of epicyclic oscillation about a nearly fixed guiding radius, and should be sampled from the same number density at this galactic radius.

\begin{figure*}
\centerline{ \epsfig{file=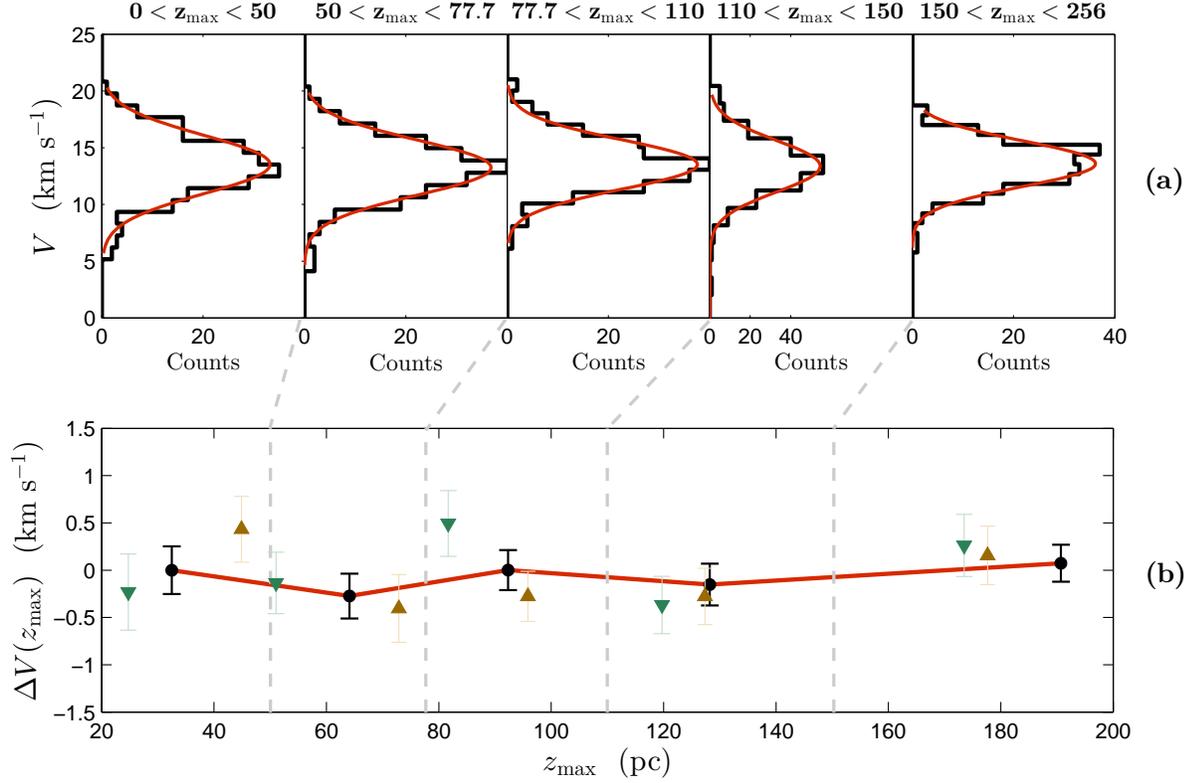,angle=0,width=0.95\textwidth}
} \caption{The average V-component deviation of the selected stars as a function of the amplitude of their vertical motion (with respect to the galactic plane). Panel {\bf(a)} shows the V distributions for the 5 $z_\mathrm{max}$ bins. Panel {\bf(b)} shows the resulting deviation of each bin's mean V from the first bin. The resulting curve is centered around zero, and shows no appreciable trend. Specifically, no lag is evident. The black errorbars are the statistical standard error of the mean calculated for each bin. The upwards (and downwards) facing triangles denote the drift velocity of the subpopulation of stars moving away (towards) the plane.}
\label{fig:AverageV} 
\end{figure*}

For this purpose we restrict our sample to stars with galactic angular momentum in the range of  $\Delta L_z = 1.5 \times 10^4 ~\mathrm{pc}~\mathrm{km}~\mathrm{s}^{-1}$ around $L_z \approx 1.85 \times 10^6 ~\mathrm{pc}~\mathrm{km}~\mathrm{s}^{-1}$. This value was chosen to avoid known streams, and samples 1109 stars in the solar neighborhood with $\left< V \right> \approx V_\odot$. The remaining kinematic distributions are therefore completely Gaussian and unaffected by any streams or an asymmetric drift. The range of $\Delta L_z$ was chosen such as to maximize the number of stars in the sample without affecting its inherent standard deviation (which is on the order of $\sigma_V \approx 2.5 ~\mathrm{km}~\mathrm{s}^{-1}$), in other words, $\Delta L_z \lesssim R \sigma_V$. We note that even without this extra precaution we obtain virtually the same constraints on MOND. This implies that the possible systematic errors from a biased population are not important in the present dataset, and that the reduced noise in this subset roughly cancels the smaller sample size. Though, it should be mentioned that the subset appears to have more Gaussian distributions.
 
We first calculate $z_\mathrm{max}$ for each star in our sample using equation \eqref{zmax} under the mentioned assumption of $\tau_0 = 64 ~\text{Myr}$. We divide the sample obeying $z_\mathrm{max}<256 ~\text{pc}$ into 5 consecutive $z_\mathrm{max}$ bins containing an equal number of stars (209 each). For each bin we then calculated the mean rotational velocity $V$ by fitting the Gaussian distribution, and plotted the results as a function of $z_\mathrm{max}$ (taken as the mean $z_\mathrm{max}$ for the population of each bin). Figure \ref{fig:AverageV} shows the resulting plot of $\Delta V(z_\mathrm{max})$ (normalized such that $\Delta V=0$ for the first bin), as well as the underlying velocity histograms for each bin. Since the distributions are very nearly Gaussian, using the arithmetic mean instead of the guassian fit results in nearly identical results.

The figure also depicts the drift velocity for the subpopulation of stars moving away and towards the galactic plane. A significant discrepancy between the two could arise if the sample of stars have a non-negligible contamination of stars which haven't been equilibrated, or if the sample is biased by the inclusion of a prominent stellar stream / moving group. The fact that no significant difference is evident between the two subsamples indicates that our population is not affected by such contamination.

When older stellar populations were included in the analysis by removing the color cuts and when the restriction on $L_z$ was lifted, the $V$ distributions were far from Gaussian and showed a prominent asymmetric drift. In this case, the $\Delta V(z_\mathrm{max})$ (calculated this time as the arithmetic mean of each bin) exhibited a negative trend of order $\sim 2 ~\mathrm{km}~\mathrm{s}^{-1}$ at $z_\mathrm{max} \sim 200 ~\mathrm{pc}$. This behavior is as expected from a contamination of the sample by an asymmetric drift, which is accentuated once older stars are introduced into the sample (older stellar populations exhibit larger epicyclic motions and can originate from smaller galactic guiding radii). Older halo and thick disk stars, which may or may not be bound to the galactic plane and are therefore not well described by our model can also contaminate the sample in this case. Although halo stars may only amount to $\sim 0.15 \%$ of the thin disk population, stars associated with the thick disk can contaminate our sample more significantly. As shown by \cite{Girard+06,MoniBidin+12}, these stars exhibit a substantial shear of $20-30 ~\mathrm{km} ~\mathrm{s}^{-1}$ at the galactic plane, and their number density is a non-negligible $~10\%$ of thin disk stars. Indeed, we find a clear indication of this thick disk population at $V \approx -25 ~\mathrm{km} ~\mathrm{s}^{-1}$ when the $L_z$ cut is removed.

\subsection{Constraining MOND parameters with the observational results}
\label{sec:MONDconstraints}

Using the observational results obtained in the previous section along with the theoretical expectation for a MONDian drift, we can constrain the MOND parameter space. The data analysis as summarized in figure \ref{fig:AverageV} is consistent with a flat $\Delta V$ curve. As such, and because the velocity drift is monotonic in $z_\mathrm{max}$, we can restrict any velocity drift at $z_\mathrm{max} = 190.7 ~\mathrm{pc}$ to be no more than $0.375~\mathrm{km}~\mathrm{s}^{-1}~ (0.823~\mathrm{km}~\mathrm{s}^{-1}$) at $1\sigma\ (2\sigma) $ confidence levels respectively (which is the value of the furthest point in figure \ref{fig:AverageV} including uncertainties for the initial, ``zero" $z_\mathrm{max}$ point).

In the following analysis we make use of eq.\ \ref{deltaV_MOND_2}, taking the values of the Oort constants as $A=14.82 \pm 0.84 ~\mathrm{km} ~\mathrm{s}^{-1} ~\mathrm{kpc}^{-1}, B=-12.37 \pm 0.64 ~\text{km} ~\text{s}^{-1} ~\text{kpc}^{-1}$ \citep{OortConstants}. We  also considered that the observed vertical oscillation frequency assuming Newtonian dynamics is $\omega_0=2\pi/64 ~\text{Myr}^{-1}$ \citep{Shaviv2014}. Note that a longer period would scale down the vertical accelerations and with it the constraints placed on MOND, inversely proportional to the fourth power of the period.

Additionally, for the solar radius we take $R_0 = 8 ~\text{kpc}$ \citep{Reid1993}. The galactic rotation velocity is then given as $V_0 = R_0(A-B) = 217.52 ~\text{km} ~\text{s}^{-1}$. For the vertical falloff of the radial field a typical order of magnitude value may be used, as given by eq.\ \ref{verticalFalloff}. This value is consistent with the value obtained by analysis of the galactic mass model of \cite{DehnenBinney1998} `Model1' (this model is discussed more in \S\ref{sec:realistic}). We are are left with two parameters we wish to constrain, $a_0$ and $\zeta$, as well as the choice of the interpolating function when evaluating eq.\ \ref{deltaV_MOND_2} at a given $z_\mathrm{max}$. The velocity drift can always be decreased by taking smaller values for the MOND acceleration $a_0$, because in the limit where $a_0 \rightarrow 0$ we revert back to a Newtonian theory. Similarly the velocity drift can be decreased by taking a smaller nonlinearity scale height $\zeta$, effectively reducing the average vertical acceleration.

We thus leave $a_0$ and $\zeta$ as free parameters, and for a given interpolating function $\mu(x)$ relate the two by solving numerically the constraint:
\begin{equation}\label{deltaVConstraint}
\Delta V\left[a_0,\zeta\right]\left(190.7 \text{pc} \right) < 0.073 \pm 0.448 ~\mathrm{km}~\mathrm{s}^{-1}
\end{equation}

\begin{figure*}
\centering
\begin{subfigure}[]
{\label{fig:a_0_zeta_a} \epsfig{file=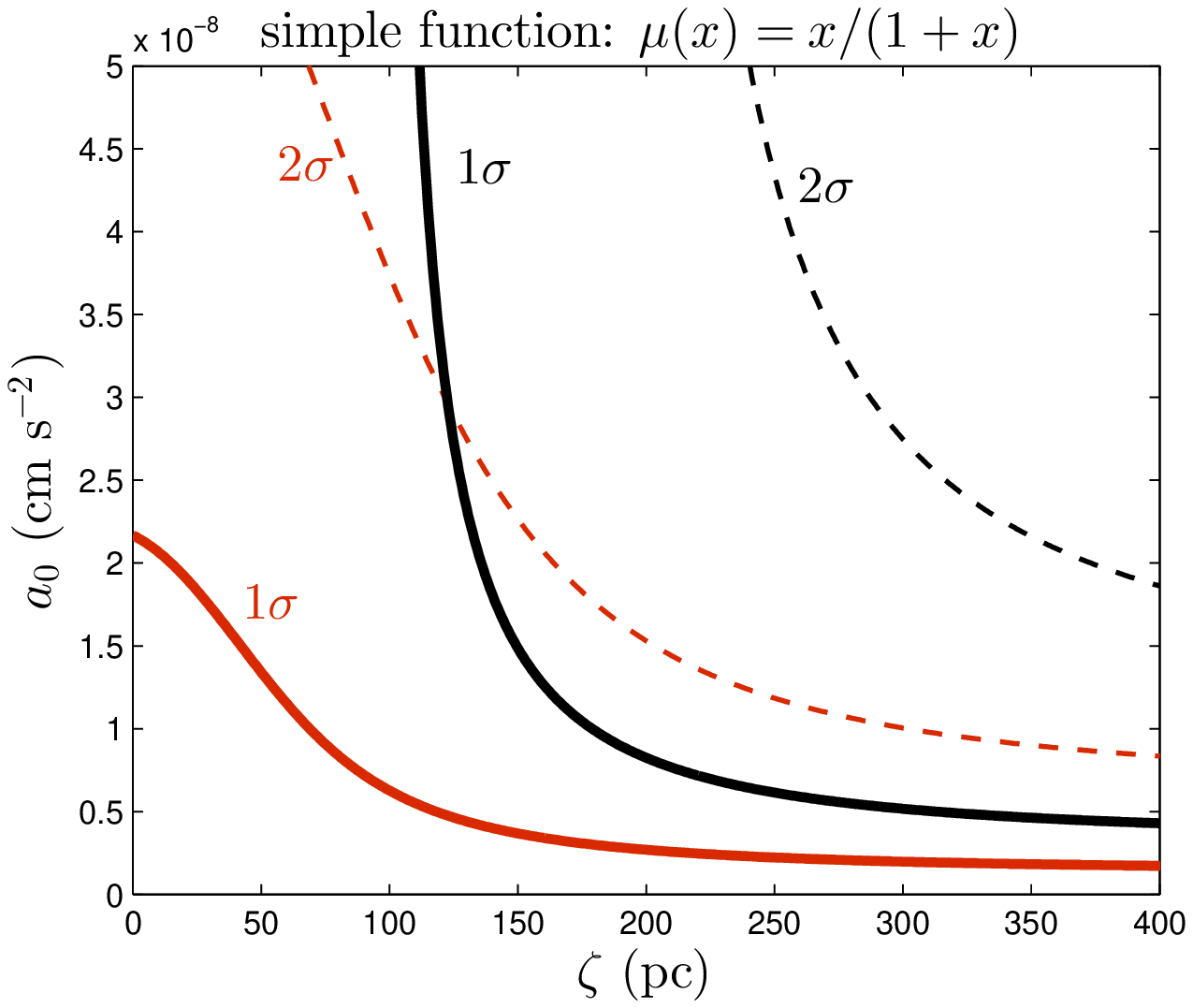,angle=0,width=0.45\textwidth} } 
\end{subfigure}
~
\begin{subfigure}[] 
{\label{fig:a_0_zeta_b} \epsfig{file=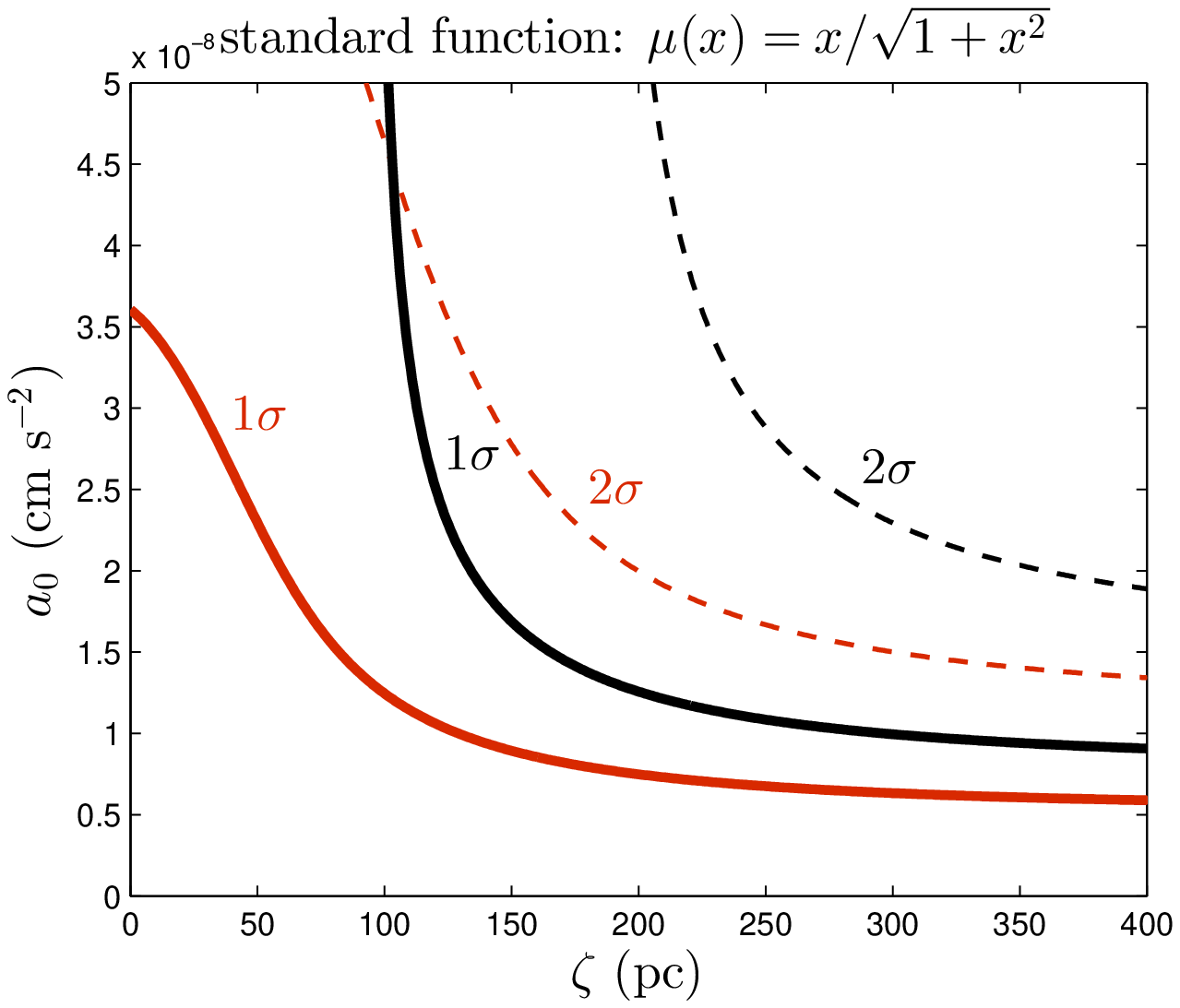,angle=0,width=0.45\textwidth} } 
\end{subfigure}
\caption{Constraints on the $(a_0,\zeta)$ parameter space for any MONDian theory. The constraint is found by evaluating eq.\ \ref{deltaV_MOND_2} and requiring that $\Delta V\left[a_0,\zeta\right]\left(190.7 ~\text{pc} \right) < 0.073 \pm 0.448 ~\mathrm{km}~\mathrm{s}^{-1}$ to be consistent with observations (as presented in \S\ref{sec:HipparcosApp}). Figure \ref{fig:a_0_zeta_a} displays the evaluated parameter space for the `simple' interpolating function, whereas figure \ref{fig:a_0_zeta_b} displays results for the `standard' interpolating function.
The regions to the right of the black solid/dashed curve in both figures illustrate the excluded regions to within one/two standard deviations (respectively), and taking a conservative lower bound for the vertical falloff of the radial gravitational field, $\partial_z^2 f_r = 0$. Similarly, the regions to the right (and above) the red solid/dashed curves show the $1/2 \sigma$ parameter space additionally excluded by taking a more realistic constraint of $\partial_z^2 f_r = 10^{-6} ~\text{Myr}^{-2} ~\text{pc}^{-1}$.} \label{fig:a_0_zeta}
\end{figure*}

We solve eq.\ \ref{deltaVConstraint} for the two commonly used interpolating functions, in each case giving both a loose constraint by taking $\partial_z ^2 f_r = 0$, and a more stringent constraint by choosing a realistic value of $\partial_z ^2 f_r = 10^{-6} ~\text{Myr}^{-2} ~\text{pc}^{-1}$. The results are plotted in Fig.\ \ref{fig:a_0_zeta}
for the standard and simple interpolating functions.
The regions to the right of the black solid (dashed) curves in both figures illustrate the excluded regions to within one (two) standard deviations (respectively), for the loose constraint (i.e. taking $\partial_z^2 f_r = 0$). Similarly, the regions to the upper-right of the red solid (dashed) curves show the $1\sigma$ ($2\sigma$) parameter space excluded by taking the more realistic estimate for $\partial_z^2 f_r$.

It is worth noting at this point that the parameter $\zeta$ is essentially the vertical scale height for the galactic disk. Traditional mass models \citep[for e.g.][]{DehnenBinney1998,BissantzGerhard2002} take this parameter as about $185 ~\text{pc}$ for the thin disk (with which we are dealing). Additionally, we can use the observation of \cite{Bahcall1985} that the harmonic period of vertical oscillations differs by about $15\%$ (possibly less) from the actual period at $z_\mathrm{max}=100 ~\text{pc}$. In our case the actual oscillation period is dependent on the scale height $\zeta$ through eq.\ \ref{eq:nonlinearz}. By plugging in the constraint given by \cite{Bahcall1985} we conclude once again that a consistent value for $\zeta$ should be about $176 ~\text{pc}$ (or more).

From figures \ref{fig:a_0_zeta_a},\ref{fig:a_0_zeta_b}, we notice that for the accepted value of the MONDian acceleration $a_0 = 1.2 \times 10^{-8} ~\text{cm}~\text{s}^{-2}$ \citep{Begeman1991}, the conservative lower bound restricts $\zeta \leq 164 ~\text{pc}$ at $1\sigma$ for the `simple' function, a value which is marginally inconsistent. The more stringent constraint on the other hand yields $\zeta \leq 57 ~\text{pc}$ within one standard deviation. The conservative constraint on the `standard' function is similar yet less limiting, because the asymptotic y-axis value of the solid curve lies higher, and gives $\zeta \leq 213 ~\text{pc}$ for $a_0 = 1.2 \times 10^{-8} ~\text{cm}~\text{s}^{-2}$, which is marginally consistent. Still, if we take into account that larger values of $a_0$ (as much as $2-3 \times 10^{-8} ~\text{cm}~\text{s}^{-2}$, \cite{FamaeyBinney2005}) are needed to fit the Milky Way rotation curve with this interpolating function, the conservative constraint will permit values of $\zeta\lesssim 130 \text{pc}$, which are below the expected value of the disk scale-height. The tighter constraint once again restricts $\zeta$ to values that are inconsistent with the $1\sigma$ constraint and may plausibly be ruled out.

Also note that we have not included the affect of $\zeta$ in the analysis of the Hipparcos data (described in \S\ \ref{sec:HipparcosApp}), where we calculated $z_\mathrm{max}$ under the assumption of a perfectly harmonic potential. Qualitatively, introducing the non-linearity in $f_z$ would increase all calculated $z_\mathrm{max}$ in fig.\ \ref{fig:AverageV} and lead to tighter constraints. 
For our specific choice of interpolation between linear and constant field (eq.\ \ref{eq:nonlinearz}), this increase in $z_\mathrm{max}$ will not compensate for the reduction of 
the average vertical acceleration.
Therefore in the limit $\zeta \rightarrow 0$ MOND can always be ``saved" (assuming $\partial_z^2 f_r = 0$).
Nonetheless, we would expect a slight shift of all curves in fig.\ \ref{fig:a_0_zeta} to the left. In this sense we have underestimated our constraints on the MOND parameter space.

\begin{figure*}
\centering
\begin{subfigure}[]
{ \epsfig{file=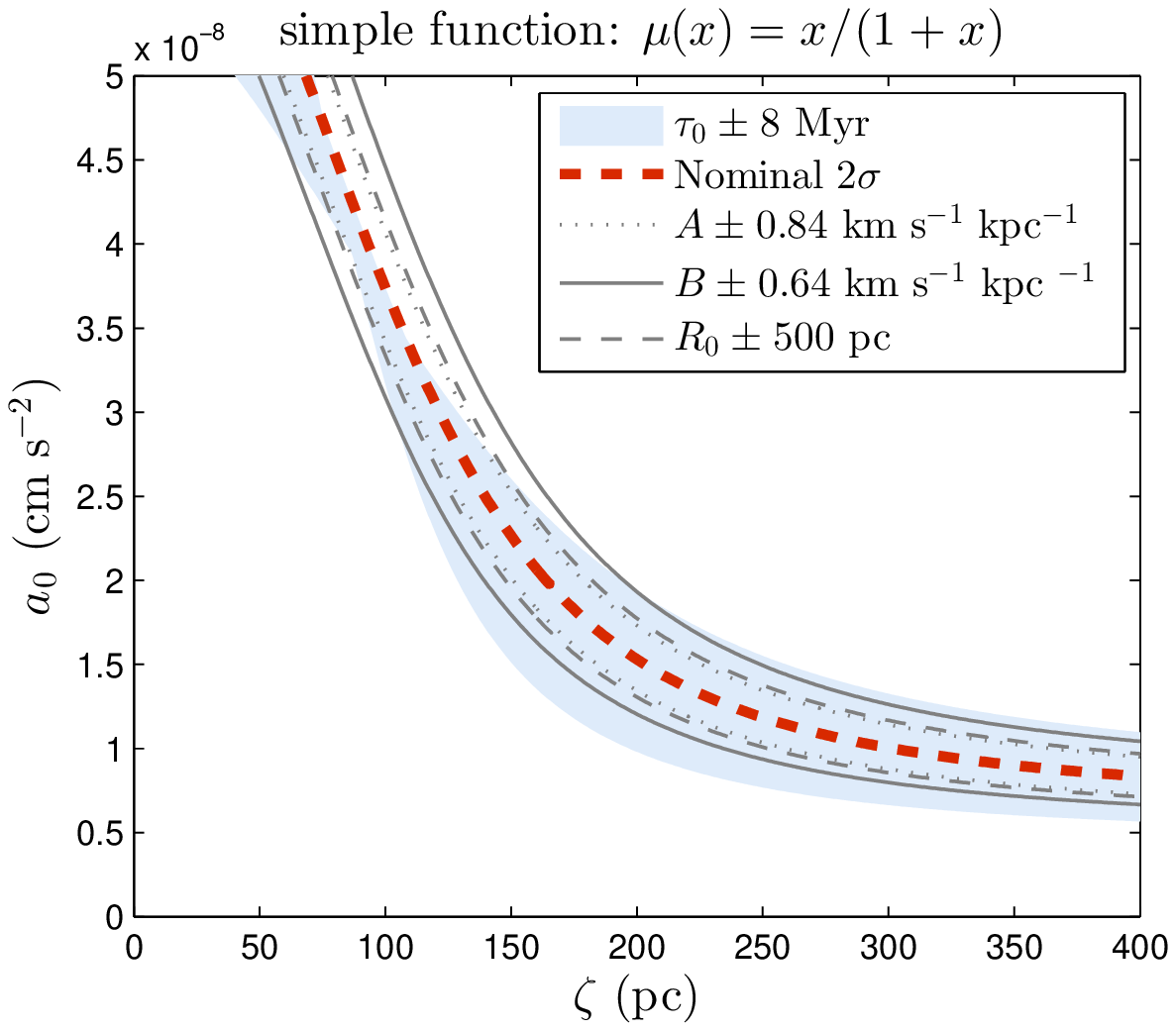,angle=0,width=0.45\textwidth} } 
\end{subfigure}
~
\begin{subfigure}[] 
{ \epsfig{file=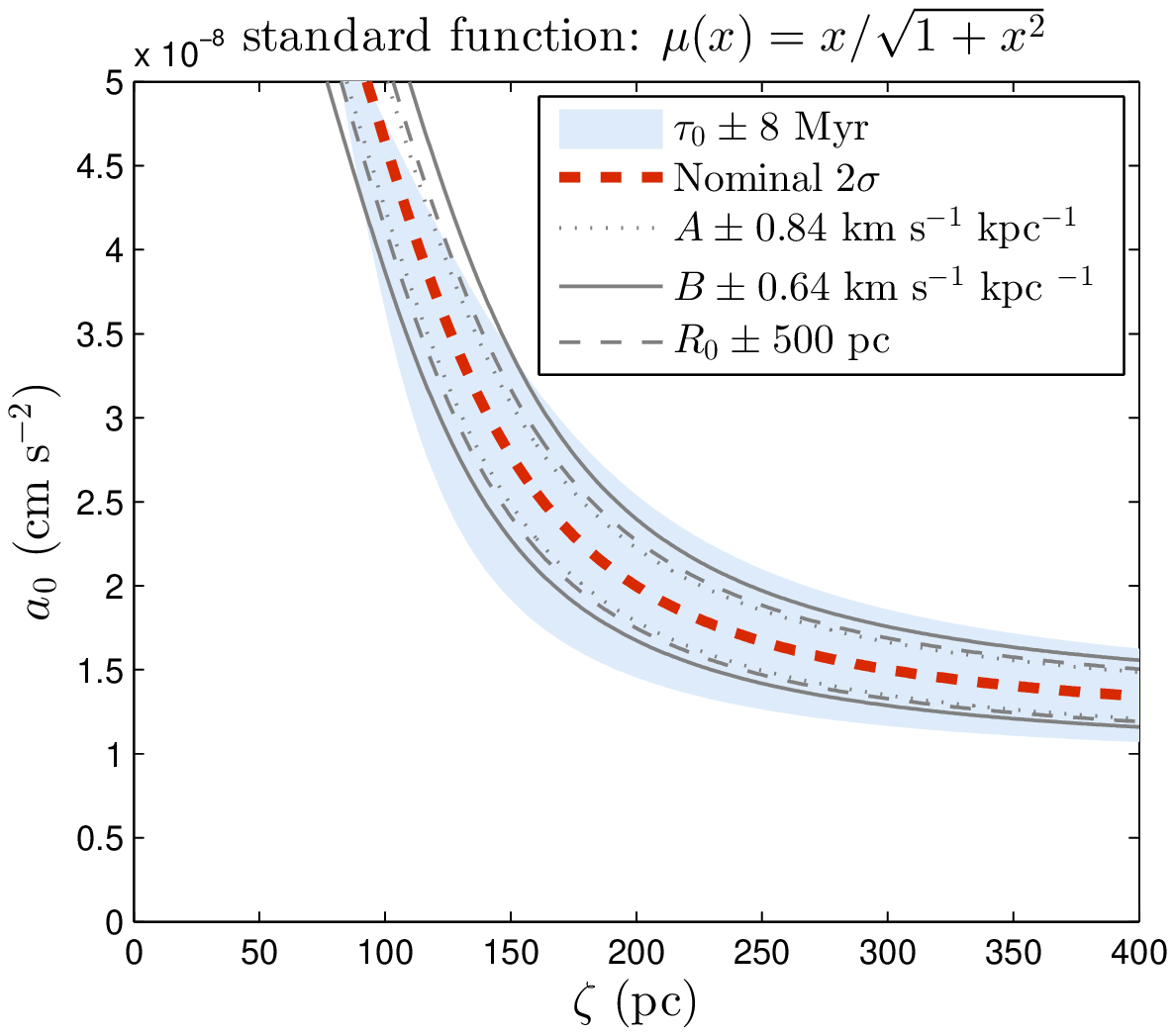,angle=0,width=0.45\textwidth} } 
\end{subfigure}
\caption{Change of the ``realistic" $2\sigma$ curves in the $(a_0,\zeta)$ parameter space under variation of the assumed physical variables within their plausible uncertainties. While the Oort constants $A$ and $B$ and the solar galactocentric radius $R_0$ only modestly alter the nominal curve, the larger deviation in $\tau_0$ affects the result more significantly.} \label{fig:a_0_zeta_Uncertainties}
\end{figure*}

Finally, we show in figure \ref{fig:a_0_zeta_Uncertainties} the change of the $2 \sigma$ realistic curves in the $(a_0,\zeta)$ plane under variation of the assumed observational parameters $R_0, A, B, \omega_0$. This provides a measure for considering the effect of the uncertainties associated with these parameters. For the Oort constants, we take the stated uncertainties of \cite{OortConstants}. Based on different studies, the solar galactocentric radius tends to vary around $8$ kpc, but is almost certainly in the range of $7.5-8.5$ kpc. We therefore take the variation in $R_0$ as $\pm 500$ pc. Finally, we allowed the vertical oscillation period to vary significantly, by $\pm 8$ Myr. This parameter is probably the most important for the MONDian drag effect, but there is also less of a consensus regarding its value which is why we permit such large fluctuations.

Each of the curves in figure \ref{fig:a_0_zeta_Uncertainties} was obtained by changing one of the parameters, while keeping the others fixed at their nominal values. It is clear that the Oort constants and galactocentric radius only modestly alter the nominal curve, so that the analysis is relatively robust for these parameters. As expected, the vertical oscillation period $\tau_0$ has a larger affect on the results. Even so, the impact of decreasing the period is less substantial than one would naively expect from plugging in larger $\omega_0$ in Eq.\ \ref{deltaV_MOND_2}. This is due to a proportional increase in all $z_\mathrm{max}$ values obtained by Eq.\ \ref{zmax}, and in particular that of the furthest bin in figure \ref{fig:AverageV}. We thus obtain tighter observational constraints which somewhat compensate for the larger $\omega_0$.

 Since we ``rule out" MOND at  only $\lesssim 2 \sigma$  (for the reasonable galactic potential), the analysis with the present dataset can only be considered as an indication, and a proof of concept for the method.

\section{Discussion}
\label{sec:discussion}

We have shown both theoretically and numerically that a rotational velocity drift should be associated with stars oscillating perpendicular to the galactic plane, and should increase in magnitude with the maximal vertical extent of the star's motion. This effect is inherently different under the assumption of MONDian versus Newtonian dynamics, and we have shown that the MONDian velocity drift is significantly larger than its Newtonian counterpart. Although our results were obtained under the QuMOND formulation of the modified dynamics, alternative MONDian theories differ from this formulation (in the non-relativistic limit) only by a curl field such as ${\bf \sigma}$. Since we have found that the main contribution to the MONDian drift arises from the algebraic part of relation \ref{eq:MOND_field}, we expect that alternative formulations will yield very similar results. In particular, eq. \ref{deltaV_MOND_2} excluding its smaller curl field contributions can be obtained exactly and directly from the non-linear Poisson formulation of MOND.

Previous works by \cite{StubbsGarg2005} and \cite{Nipoti2007} have been motivated by the same concept of interplay between vertical and radial dynamics in MOND, yet differ from our current work in several aspects. Specifically, neither of these works manage to quantitatively constrain MOND with current observational data. \cite{StubbsGarg2005} method is also somewhat limited in its applicability, as it specifically assumes an exponential disk model with a non-varying scale height, and relies on data measured over a large range of galactic radii (throughout which this assumption most likely does not hold). Our present method on the other hand, relies only on local kinematic observables which can be better constrained. Being also well aware of the limitations of the harmonic approximation, we explicitly introduce and allow deviation from this model by proper choice of the free parameter $\zeta$, which we carry throughout.

By using complete kinematic data for stars in the local galactic neighborhood obtained from the XHIP catalog, we have shown that observations are consistent with no apparent velocity drift up to $z_\mathrm{max}=190 ~\text{pc}$ and within our confidence bounds of $\pm 0.448 ~\text{km}~\text{s}^{-1}$. This fact allows us to constrain the $(a_0,\zeta)$, and $\mu(x)$ parameter space by requiring that they will be consistent with this observation. We use our analytic expression (lower-bound) for the velocity drift (eq.\ \ref{deltaV_MOND_2}) as a function of these parameters and conclude that current accepted values for $a_0$ and the disk scale-height are marginally inconsistent with our presented observational constraint. By assuming a more realistic value for the vertical falloff of the radial gravitational field ($\partial_z^2 f_r = 10^{-6} \text{Myr}^{-2} \text{pc}^{-1}$) we find that reasonable values for these two parameters can be ruled out at $68\%$ confidence bounds, and conclude that present MOND theories may suffer from inconsistencies. Since this initial analysis yields only weakly significant constraints, it serves primarily as an indication of such possible MONDian inconsistencies, and should more importantly be interpreted as a proof of concept for our method.

We expect that even the conservative constraint (neglecting $\partial_z^2 f_r$) may be improved in the near future with the GAIA mission \citep{GAIA2001}, by minimizing the error bounds on $\Delta V$ and possibly increasing the maximal observed $z_\mathrm{max}$. Even a decrease of these error bounds by a mere factor of two will significantly impact the results as the theory is highly non-linear. By increasing the sample size and detection horizon, we may also be able to constrain the velocity drift at larger $z_\mathrm{max}$ than present, which will possibly constrain the $(a_0,\zeta)$ parameter space even further. It is important to note that our method was developed as a series expansion about the galactic plane, and may thus break down at very large vertical extents. Nonetheless, even a modest increase in $z_\mathrm{max}$ to $\sim 300$ pc (which is likely still well described by our method, especially since we take into account the non-linearity of the potential at larger heights via the parameter $\zeta$) would tighten the constraints on MOND significantly.

An interesting by-product result of our analysis is an upper bound on $\partial_z^2 f_r$ assuming normal Newtonian dynamics. This bound is easily derived by taking the Newtonian limit of eq.\ \ref{deltaVConstraint}, i.e. $\mu=1, \mu^{\backprime}=0$, and solving for $\partial_z^2 f_r$. Using this procedure we find: $\left( \partial_z^2 f_r \right)_0 < 2.1 \times 10^{-6} ~\text{Myr}^{-2} ~\text{pc}^{-1}~(4.7 \times 10^{-6} ~\text{Myr}^{-2} ~\text{pc}^{-1})$ at $1\sigma$ ($2 \sigma$) confidence level respectively. This result may be useful in constraining Newtonian galactic mass models, but lies outside the scope of our present work.

\section*{Acknowledgements}
NJS gratefully acknowledges the support of the IBM Einstein Fellowship. This research project was supported by the
I-CORE Program of the Planning and Budgeting Committee and the Israel Science Foundation (center 1829/12).

\bibliography{VerticalBibliography2}

\end{document}